 \documentclass[%
reprint,
amsmath,amssymb,
aps, 
]{revtex4-2}

\usepackage{amsmath,amssymb}
\usepackage[dvipdfmx]{graphicx}
\usepackage{hyperref}
\usepackage{physics}
\usepackage{color,soul}

\newcommand{\pd}{{\partial}}
\newcommand{\ee}{{\mathrm{e}}}
\newcommand{\ri}{{\mathrm{i}}}

\begin{document}

\title{Numerical search for states with constant enstrophy flux 
  over finite time intervals \\
  in two-dimensional turbulence}
\author{Kyo Yoshida}
\affiliation{%
Department of Physics, Institute of Pure and Applied Sciences, University of Tsukuba
}
\date{\today}

\begin{abstract}
  An ensemble model of turbulence based on states with constant flux in wavenumber space was proposed in [K. Yoshida, \emph{Phys. Rev. E}, {\bf 106}, 045106 (2022)].  The justification of this ensemble model relies on the conjecture that almost all states with constant flux correspond to turbulence states.  To verify this conjecture, a numerical search for states with constant enstrophy flux in wavenumber space over finite time intervals in two-dimension turbulence is conducted using a Monte Carlo method.  Properties of the obtained states, such as energy spectrum and spectra related to fourth-order moments, are examined and compared with those of turbulence states obtained from direct numerical simulations.  The dependence of the obtained states on the time interval and the initial conditions used in the numerical search is also discussed.  

\end{abstract}

\maketitle

\section{Introduction}
\label{sec:introduction}
The motions of viscous fluids can be modeled by the Navier–Stokes (NS) equations. Although governed by the deterministic equations, fluid motions become seemingly irregular when viscosity is small. We refer to such states of flow as turbulence. 
Turbulence is a nonequilibrium state in the sense that there is a macroscopic dissipation of energy by viscosity, and continuous injection of energy by external forces is required to maintain it. Therefore, equilibrium statistical mechanics, which is established based on ensemble models such as microcanonical and canonical ensembles, cannot be directly applied to turbulence.

Many attempts have been made to introduce appropriate ensembles for turbulence from various perspectives, including rigorous mathematics~\cite{Bedrossian2022batchelor}, periodic orbit theory~\cite{KawaharaKida2001}, and field-theoretic formalisms~\cite{MartinSiggiaRose1973, Janssen1976, deDominicis1976, CanetDelamotteWschebor2016}. In statistical closure approaches (see, e.g., \cite{Kraichnan1965,Kaneda1981}), low-order moments are analyzed under certain assumptions without explicitly specifying an ensemble. See Ref.~\cite{Zhou2021} for a comprehensive review of statistical closure approaches. Despite these efforts, it may be said that there is no established statistical theory of turbulence comparable to ensemble models in equilibrium statistical mechanics to date.

An ensemble model for turbulence based on states with constant flux in wavenumber space was proposed in Ref.~\cite{Yoshida2022}, hereafter referred to as Y22. The model incorporates the phenomenology of the energy cascade~\cite{inRichardson1922, Kolmogorov1941a} at the level of its construction. In wavevector space, the energy cascade is expressed as $\Phi_k = \epsilon$, where $\Phi_k$ is the energy flux from the small-wavenumber region $\{\vb*{p}|\ |\vb*{p}|<k\}$ to the large-wavenumber region $\{\vb*{p}|\ |\vb*{p}|\ge k\}$ due to nonlinear interactions in the NS equations, and $\epsilon$ is the energy dissipation rate, independent of $k$. The ensemble consists of all states such that $\Phi_k = \epsilon$ for $k_0 \le k \le k_{\mathrm{max}}$ is maintained under inviscid dynamical evolution over a time interval $T$. 

\begin{figure}[t]
\includegraphics[width=0.45\textwidth]{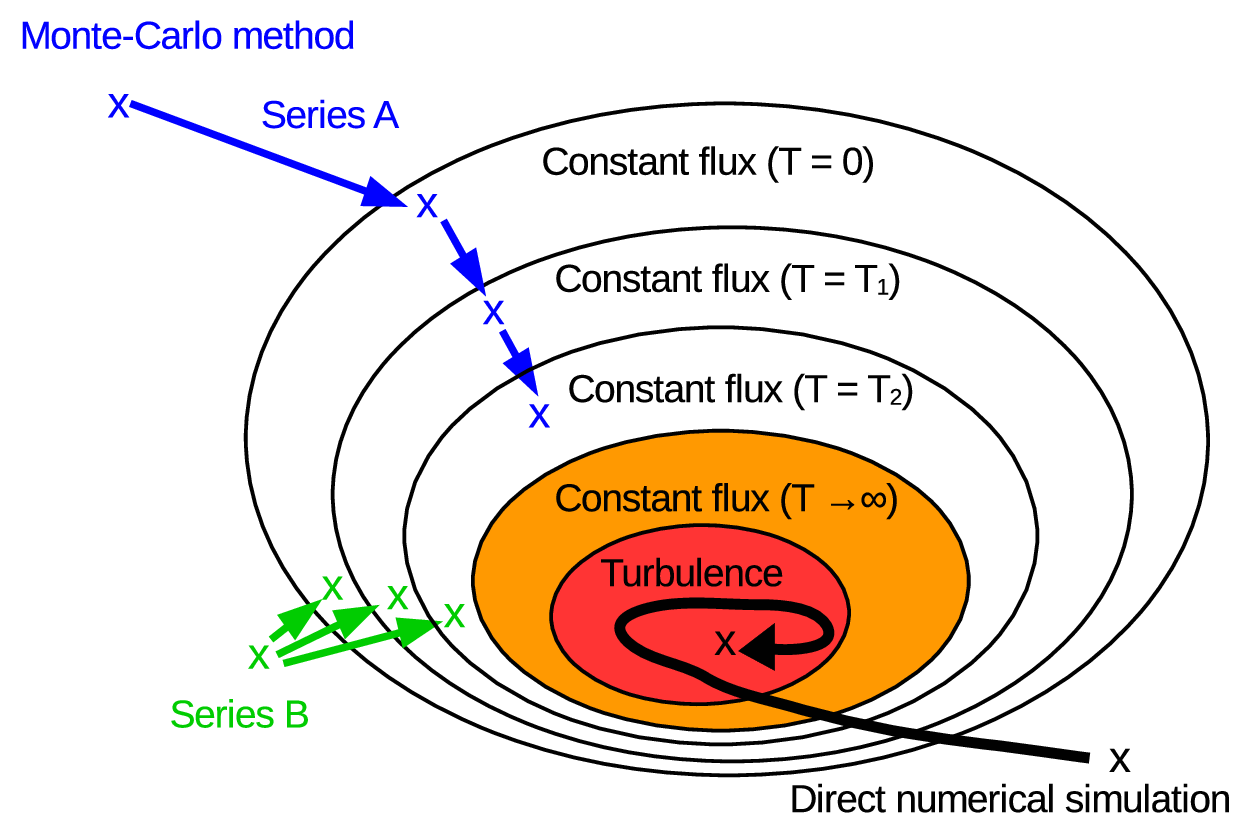}
\caption{Conceptual diagram of ensembles of states with constant flux over time intervals and the ensemble of turbulence states.  Numerical searches for states with constant flux using a Monte Carlo method are depicted schematically. See Sec.~\ref{sec:setup} for explanations of Series A and B.}
\label{fig:ensemble}
\end{figure}

In many direct numerical simulations (DNS) of forced NS equations in periodic boundary boxes, it is observed that the energy flux relaxes under dynamical evolution to a quasi-constant $\Phi_k \sim \epsilon$ in the inertial range, although the inertial range is limited due to computational resources.  
  In general, $\epsilon$ depends on time with a characteristic time scale, say $\tau_\epsilon$, reflecting the nature of the large-scale forcing. Since the characteristic time scale of turbulence at wavenumber $k$ may be estimated as $\tau(k):=\epsilon^{-1/3}k^{-2/3}$, we may expect quasi-stationarity of the flux $\Phi_k$ for sufficiently large $k$ in the inertial range such that $\tau(k)\ll \tau_\epsilon$. (See Ref.~{\cite{BosRubinstein2017}} for corrections arising from the nonstationarity of $\epsilon$.) It is indeed observed in DNS that the fluctuation of $\Phi_k$ about $\epsilon$ decreases as $k$ increases, indicating that $\Phi_k=\epsilon$ holds for sufficiently large $k$ with a certain degree of accuracy in an instantaneous turbulent state, and that this relation persists over a finite period of time (see, e.g., Figs.~2 and 3 of Ref.~{\cite{Ishiharaetal2016}}).

Let us define turbulence states as those in the attractor of NS dynamics. We may then expect that, at sufficiently large Reynolds numbers, turbulence states are states with constant energy flux in some wide wavenumber range $k_0 \le k \le k_{\mathrm{max}}$ over some long time interval $T$. In other words, the ensemble of turbulence states is included in the ensemble of states with constant energy flux. See Fig.~\ref{fig:ensemble} for a conceptual diagram.  

Here, we propose the following conjecture. The ensemble of states with constant flux over a sufficiently long time interval coincides with the ensemble of turbulence states, except for atypical states that occupy an extremely small measure in comparison to the whole of either ensemble. If this conjecture holds, then we can compute statistical characteristics of turbulence by taking averages of suitable quantities over the ensemble of states with constant flux, although such computation would be quite difficult and methods would need to be developed to obtain approximate values.

The conjecture itself is yet a rough statement, and detailed conditions for both the states with constant flux and the turbulence states should be considered for rigorous mathematical discussions. Instead of a mathematically rigorous approach, an alternative numerical approach was attempted in Y22 and will be pursued in the present study to verify the plausibility of the conjecture.

In Y22, a Monte Carlo method was employed to numerically search for states with quasi-constant enstrophy flux in a two-dimensional incompressible fluid system.  Focusing on two-dimensional turbulence instead of three-dimensional was due to the preliminary nature of the work and limited computational resources.  The error function to minimize in the Monte Carlo method was set to search for states with constant enstrophy flux over the whole wavenumber range of the simulation with $T=0$, that is, instantaneously in time. See Sec.~\ref{sec:numerical_search} for details of the Monte Carlo method. It was found that even though $T=0$, the states with quasi-constant enstrophy flux obtained via the Monte Carlo method resembled turbulence states in terms of their energy spectra. This result provides positive support for the conjecture. However, the vorticity field patterns of the states with constant flux and the turbulence states were apparently different.

In this paper, we perform numerical searches for states with constant enstrophy flux over time intervals $T>0$ in order to further verify the conjecture. Note that the ensemble of states with constant flux over a time interval $T_1$ is a subset of that with time interval $T_0$ when $T_1 > T_0$ 
because states with constant flux over $T_1$ must maintain that constant flux throughout the additional interval $T_1-T_0$.
The ensemble becomes smaller as the time interval $T$ increases (see Fig.~\ref{fig:ensemble}). If the conjecture is correct, then the ensemble of states with constant flux over a sufficiently long time interval almost coincides with the ensemble of turbulence states, so that the states obtained in the numerical searches would resemble turbulence states not only in terms of energy spectra but also in other quantities, e.g., spectra related to fourth-order moments, which will be analyzed in the present paper.

This paper is organized as follows. The ensemble model of turbulence based on states of constant flux and the method of numerical search for such states are reviewed in Secs.~\ref{sec:ensemble_model} and \ref{sec:numerical_search}, respectively. The setup of the present numerical search is given in Sec.~\ref{sec:setup}, and the results are presented in Sec.~\ref{sec:results}. Discussions of the results are provided in Sec.~\ref{sec:discussions}. Some data of turbulence states obtained in DNS are shown in Appendix 
for comparison with those of the obtained states in the numerical searches.

\section{Ensemble model}
\label{sec:ensemble_model}

We consider an incompressible fluid in a two-dimensional domain 
$[0,L]^2$ with periodic boundary conditions. 
A state of the fluid is specified by an incompressible velocity 
vector field $\vb*{u}(\vb*{x})$ or the vorticity field 
$\omega(\vb*{x})=\epsilon_{jl}\pd_j u_l(\vb*{x})$, where 
$\epsilon_{jl}$ is the antisymmetric tensor, 
$\epsilon_{12}=-\epsilon_{21}=1$ and $\epsilon_{11}=\epsilon_{22}=0$.  
Here and hereafter, $a_j$ denotes the $j$-th component of the vector $\vb*{a}$, 
and summation over repeated component indices is assumed.  
Let $\omega_{\vb*{k}}:=(2\pi)^{-2} \int_{[0,L]^2}\dd{\vb*{x}} 
\ee^{-\ri \vb*{k}\cdot\vb*{x}}\omega(\vb*{x}) 
\ (\vb*{k}\in\mathcal{K})$ denote the Fourier 
coefficients of the vorticity field, 
where $\mathcal{K}$ is the set of wavevectors 
$\mathcal{K}:=\{(k_1,k_2)|k_j=m\Delta k, m\in \mathbb{Z}, 
k< k_{\max}\}-\{\vb*{0}\}$, with $k:=|\vb*{k}|$, $\Delta k:=2\pi/L$, 
and a cutoff wavenumber $k_{\max}$ is introduced. 
The reality of $\omega(\vb*{x})$ in physical space implies 
$\omega_{-\vb*{k}}=\omega_{\vb*{k}}^*$. The Fourier coefficients of 
the velocity field $\vb*{u}_{\vb*{k}}$ are related to $\omega_{\vb*{k}}$ 
as $u_{\vb*{k},j}=\epsilon_{jl}(k_l/k^2)\omega_{\vb*{k}}$.  
In the following, we symbolically denote the state by 
$\vb*{\omega}$. 

The NS equation in wavevector space is given by
\begin{equation}
\dv{t}\omega_{\vb*{k}}(t)= 
M_{\vb*{k}}\qty(\vb*{\omega}(t)) - \nu k^2 \omega_{\vb*{k}}(t) 
+ \ri \epsilon_{jl} k_j f_l(t), 
\label{eq:NSk}
\end{equation}
where the mass density of the fluid is unity, 
$\nu$ is the kinematic viscosity, 
$\vb*{f}(t)$ is the external forcing field, and 
$\vb*{M}$ is a map from a vector field to a vector field given by 
\begin{align}
M_{\vb*{k}}\qty(\vb*{\omega})&=
\sum_{\vb*{p}}^{\Delta}\sum_{\vb*{q}}^{\Delta}
\delta_{\vb*{k}-\vb*{p}-\vb*{q}}^{\Delta}
\frac{1}{2}\qty(\frac{1}{p^2}-\frac{1}{q^2})\epsilon_{jl}p_jq_l 
\omega_{\vb*{p}}\omega_{\vb*{q}}, 
\end{align}
where $\sum_{\vb*{k}}^\Delta:=\sum_{\vb*{k}\in\mathcal{K}}\qty(\Delta k)^2$, 
$\delta^{\Delta}_{\vb*{k}}=(\Delta k)^{-2}$ for $\vb*{k}=\vb*{0}$ 
and $\delta^{\Delta}_{\vb*{k}}=0$ otherwise. 

The enstrophy density per unit volume, or simply enstrophy hereafter,  
$\Omega(\vb*{\omega})$ 
is given by 
\begin{equation}
\Omega(\vb*{\omega})=\sum_{\vb*{k}}^\Delta \Omega_{\vb*{k}}(\vb*{\omega}), \quad
\Omega_{\vb*{k}}(\vb*{\omega}):=\frac{1}{2}\qty(\Delta k)^2 |\omega_{\vb*{k}}|^2, 
\end{equation}
where $\Omega_{\vb*{k}}(\vb*{\omega})$ is the enstrophy of the wavevector mode $\vb*{k}$. The energy spectrum of the state $\vb*{\omega}$ is defined by
\begin{equation}
E_k(\vb*{\omega}):=(\Delta k)^{-1}
\sum_{\substack{\vb*{p}\\(k-\Delta k/2 \le p<k+\Delta k/2)}}^{\Delta}
p^{-2}\Omega_{\vb*{p}}(\vb*{\omega}). 
\end{equation}

Hereafter, let $\vb*{\omega}(t)$ denote the solution of (\ref{eq:NSk})
with $\nu=0$, $\vb*{f}(t)=\vb{0}$, and 
initial condition $\vb*{\omega}$ at $t=0$.  
The enstrophy flux $\Phi^\Omega_{k}(\vb*{\omega})$ 
from the small-wavenumber region $\{\vb*{p}|p<k\}$ 
to the large-wavenumber region $\{\vb*{p}|p\ge k\}$ due to 
the interaction represented by $\vb*{M}$ is given by 
\begin{equation}
\Phi^\Omega_{k}(\vb*{\omega})
:=-\qty(\Delta k)^2\sum_{\vb*{p}(p<k)}^\Delta 
\Re \qty(M_{\vb*{p}}(\vb*{\omega}){\omega}_{-\vb*{p}}). 
\label{eq:Phi}
\end{equation}

An ensemble of states is specified by a probability density function 
$P(\vb*{\omega})$ satisfying $P(\vb*{\omega})\ge 0$ and 
$\int \mathcal{D}\vb*{\omega} P(\vb*{\omega})=1$, where 
$\mathcal{D}\vb*{\omega}:=\prod_{\vb*{k}\in \mathcal{K}^+} \dd{\omega_{\vb*{k}}}, 
\dd\omega_{\vb*{k}}:=\dd{\Re(\omega_{\vb*{k}})}\dd{\Im(\omega_{\vb*{k}})}$, 
$\mathcal{K}^+(\subset \mathcal{K})$ is a set of wavevectors such 
that for all $\vb*{k}\in\mathcal{K}$, either $\vb*{k}\in\mathcal{K}^+$ or $-\vb*{k}\in\mathcal{K}^+$ but not both. 
The ensemble average of a function $F(\vb*{\omega})$ is given by
$\langle F(\vb*{\omega})\rangle 
:=\int \mathcal{D}\vb*{\omega} P(\vb*{\omega}) F(\vb*{\omega})$. 

An ensemble model of turbulence based on states of constant flux in 
wavenumber space was proposed in Y22.  
For the enstrophy cascade range of two-dimensional turbulence, 
the probability density function of the model is given by 
\begin{equation}
P_\eta(\vb*{\omega}):=C \prod_{n=0}^{N_t}\prod_{m=0}^{N_k}
\delta\qty\bigg(\Phi^\Omega_{k_m}\qty\big(\vb*{\omega}(t_n))-\eta), 
\label{eq:Pfluxconst}
\end{equation}
where $\delta(x)$ is the Dirac delta function, $C$ is 
the normalization constant, 
$\eta$ is a constant corresponding to the enstrophy dissipation rate, 
$0<k_{\min}=k_0<\ldots<k_{N_k}=k_{\max}$, and $0=t_0<t_1<\ldots<t_{N_t}=T$. 
Formally, by taking the limits $N_k,N_t\to \infty$ with 
$\min_{m}(k_{m+1}-k_m), \min_{n}(t_{n+1}-t_n)\to 0$, 
one obtains an ensemble of states with constant enstrophy flux 
$\Phi^\Omega_k(\vb*{\omega}_t)=\eta$ in the wavenumber range 
$k_{\min}\le k \le k_{\max}$ and over the time interval $0\le t \le T$.  
By further taking the limits $k_{\max}\to \infty$ and $T\to \infty$, 
one obtains a stationary ensemble model of states with 
constant enstrophy flux for $k\ge k_0$.  

A possible refinement of the model would be to replace the Dirac delta function $\delta(x)$ with a function $f(x)$ that has a sharp peak at $x = 0$ and a finite variance, $\int \dd x\, f(x)\, x^2 > 0$, in order to account for flux fluctuations. However, these fluctuations are expected to decrease as the wavenumber increases and may become negligible for sufficiently large $k_0$ and in the limit $k_{\max} \to \infty$. Therefore, such a refinement is not considered at the present stage.

\section{Numerical search of states}
\label{sec:numerical_search}
If typicality applies to the present ensemble model, 
some properties of turbulence should be possessed by 
a single typical state in the ensemble without taking 
the ensemble average.  

A numerical method to search for a single state from the ensemble was 
introduced in Y22. 
A Monte Carlo (MC) method was used to minimize a certain error function of 
the state. In this paper, we consider a slightly different type of 
error function from that in Y22, defined by 
\begin{equation}
\Delta^{(T)}(\vb*{\omega}):=\frac{1}{(N_k+1)(N_t+1)}
\sum_{m=0}^{N_k}\sum_{n=0}^{N_t}
\qty({\Phi^\Omega}_{k_m}(\vb*{\omega}(t_n))-\eta)^2, 
\label{eq:Delta}
\end{equation}
where $T=t_{N_t}$. 

The MC step associated with $\vb*{k}(\in\mathcal{K}^+)$,  
which updates a given state $\vb*{\omega}$ to a new one, 
is given by the following substeps: \\

(1) Let 
\begin{equation}
\omega'_{\vb*{k}}=\omega_{\vb*{k}}\exp\qty(r \ee^{\ri\theta}),\quad
\omega'_{-\vb*{k}}=(\omega'_{\vb*{k}})^*, 
\end{equation}
and $\omega'_{\vb*{p}}=\omega_{\vb*{p}}$ for $\vb*{p}\ne \pm\vb*{k}$, 
where $r$ is a fixed parameter satisfying $0<r<1$, 
and $\theta$ is a uniform random variable on $[0,2\pi)$. \\

(2) The transition probability is defined by
\begin{equation}
T(\vb*{\omega}',\vb*{\omega})=
\begin{cases}
\min\qty(\frac{|\omega'_{\vb*{k}}|^2}{|\omega_{\vb*{k}}|^2},1)
& (\Delta^{(T)}(\vb*{\omega}')-\Delta^{(T)}(\vb*{\omega})\le 0),\\
0 
& (\Delta^{(T)}(\vb*{\omega}')-\Delta^{(T)}(\vb*{\omega})> 0).
\end{cases}
\end{equation}
Accept $\vb*{\omega}'$ as the new state of $\vb*{\omega}$ 
with probability $T(\vb*{\omega}',\vb*{\omega})$, and 
keep $\vb*{\omega}$ unchanged otherwise.  

Since the typical scale of $\omega_{\vb*{k}}$ is not known a priori, 
we set a uniform step amplitude $r$ in $(\ln \omega_{\vb*{k}})$-space. 
The transition probability $T(\vb*{\omega}', \vb*{\omega})$ corresponds to the Metropolis algorithm with a modification factor due 
to the nonuniform step in $\omega_{\vb*{k}}$-space. 

An MC cycle consists of performing MC steps associated with 
each $\vb*{k}\in\mathcal{K}^+$ once, in order of increasing $k$.  
In a run of numerical search, a given initial state 
is developed through repeated MC cycles.   
We regard that a state from the ensemble model $P_\eta(\vb*{\omega})$ 
is numerically sampled with a precision $\delta$ 
if $\Delta^{(T)}(\vb*{\omega})<\delta$ is achieved during the run.
Note that, when $\delta>0$, the inclusion relationships among the ensembles of states with quasi-constant flux over different time intervals, as shown in Fig.~{\ref{fig:ensemble}}, no longer hold in a strict sense.

\section{Set up}
\label{sec:setup}

\begin{table}[t]
  \caption{\label{tab:runs}Parameters of the numerical search runs.
    $T$ is the time interval, $\tau:=\eta^{-1/3}$ is the typical turbulence time scale, 
$c$
 is the number of MC cycles, and $\delta$ is the achieved precision. }
\begin{ruledtabular}
\begin{tabular}{llrl}
  Run &  $T/\tau$ & 
$c$
  & $\qquad \delta$\\
\hline
A0 & $0$ & $800$ & $0.267\times 10^{-4}$ \\
A1 & $0.02$  & $32$ & $0.494\times 10^{-3}$ \\
A2 & $0.04$  & $20$ & $0.111\times 10^{-2}$ \\
A3 & $0.08$  & $20$ & $0.171\times 10^{-2}$ \\
A4 & $0.125$ & $16$ & $0.340\times 10^{-2}$ \\
A5 & $0.25$  & $20$ & $0.935 \times 10^{-2}$ \\
B0 & $0$ & $160$ & $0.299\times 10^{-5}$ \\
B1 & $0.02$  & $32$ & $0.676\times 10^{-2}$ \\
B2 & $0.04$  & $64$ & $0.327\times 10^{-2}$ \\
B3 & $0.08$  & $64$ & $0.764\times 10^{-2}$ \\
B4 & $0.125$ & $40$ & $0.128\times 10^{-1}$ \\
B5 & $0.25$ & $24$ & $0.338\times 10^{-1}$ \\
\end{tabular}
\end{ruledtabular}
\end{table}

The numerical search for states with constant enstrophy flux 
for $T=0$ ($N_t=0$) was attempted in Y22. 
The initial states of these numerical searches were chosen to be 
states whose amplitudes $|\omega_{\vb*{k}}|$ are sufficiently small 
such that the energy spectrum $E_k(\vb*{\omega})$ is smaller than 
that of the enstrophy cascade range spectrum 
\begin{equation}
E_k=C_K \eta^{2/3}k^{-3}(\ln(k/k_\mathrm{b}))^{-1/3}
\label{eq:Ek_ens}
\end{equation}
with $C_K=1.81$, as estimated by 
the Lagrangian renormalized approximation (LRA)~\cite{Kaneda1987,Kaneda2007}, 
which is in good agreement with numerical simulations 
\cite{IshiharaKaneda2001}. 
It was found that the energy spectra of the states obtained in 
the numerical searches in Y22 are consistent with 
Eq.~(\ref{eq:Ek_ens}). 

In this study, we attempt numerical searches for states with 
constant enstrophy flux over time intervals $T>0$.  
In the case of $T>0$, it is necessary to compute the time evolution 
of $\vb*{\omega}(t)$ for the time interval $[0,T]$ at every MC step.  
The computational time for an MC step or an MC cycle increases almost 
linearly with $T$, and the feasible number of MC cycles is limited 
by available computational resources. For the sake of 
saving computational cost, we abandon for the MC runs with $T>0$ 
the initial condition used in Y22 (i.e., 
a state with sufficiently small amplitudes of $\omega_{\vb*{k}}$). 
Instead, we consider two series of runs, referred to as 
Series A and B, as follows. 

Let $T_j$ ($j=0,1,2,\ldots$) be an increasing series of time intervals, 
$0=T_0<T_1<T_2,\dots$.  
The error function of the $(j+1)$-th run of Series X 
($j=0,1,2,\ldots$, X=A,B), RunX$j$, is given by 
$\Delta^{(T_j)}(\vb*{\omega})$. The initial state
of RunA$0$ is given by $\omega_{\vb*{k}}=\beta^{-1/2}\exp(\ri\theta_{\vb*{k}})$, 
where $0 < \beta<\infty$ 
and $\theta_{\vb*{k}}$ are uniform random variables on $[0,2\pi)$. 
The initial state of RunA$j$ ($j\ge 1$) is the final state of RunA$(j-1)$. 
The initial state of RunB$j$ ($j\ge 0$) is given by 
$\omega_{\vb*{k}}=\pi^{-1/2}k^{-1/2}(\Delta k)^{-1}E_k^{1/2}\exp(\ri\theta_{\vb*{k}})$, 
with $E_k$ given by Eq.~(\ref{eq:Ek_ens}) and 
$\theta_{\vb*{k}}$ uniform random variables on $[0,2\pi)$. 
The schematic structure of these two series of numerical searches is shown in Fig.~\ref{fig:ensemble}. 

The number of grid points in the periodic domain $[0,L]^2$ is $N^2$. 
A Fourier spectral method with phase shifting is used for the computation 
of the nonlinear terms, and the maximum wavenumber is
$k_{\max}=(\sqrt{2} N/3) \Delta k$
, where $\Delta k=2\pi/L$. A fourth-order Runge–Kutta method is used for the dynamical evolution of the states. We set $N=512$, $\Delta k=1$ ($L=2\pi$), and $\Delta t=0.125\times 10^{-2}$ for the time step of the dynamical evolution.  The time intervals $T$ are listed in Table~\ref{tab:runs} together with the number of executed MC cycles
$c$
and the achieved precisions $\delta$ for each run.  

The values of the parameters associated with the Monte Carlo method are 
$\eta=1$, $r=0.0625$, $\beta=10^8$, $N_k=239$, $k_m=(m+0.5)\Delta k$ ($0\le m \le N_k$), and $t_n=n\Delta t$ ($0 \le n \le T_j/\Delta t$). The time intervals are 
$T_1=16\Delta t$, $T_2=32\Delta t$, $T_3=64\Delta t$, $T_4=100\Delta t$, and $T_5=200\Delta t$. These may be written in terms of the typical turbulence time scale defined by $\tau:=\eta^{-1/3}$ as $T_1=0.02\tau$, $T_2=0.04\tau$, $T_3=0.08\tau$, $T_4=0.125\tau$, and $T_5=0.25\tau$.

\section{Results}
\label{sec:results}

\begin{figure}[t]
\includegraphics[width=0.45\textwidth]{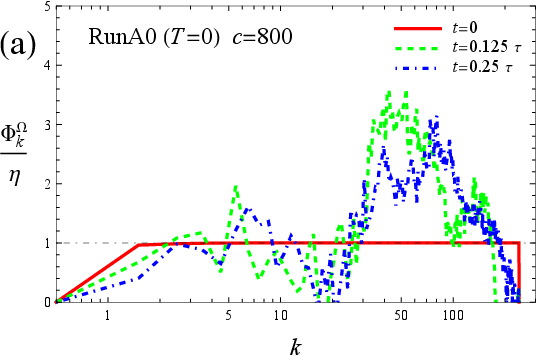}
\includegraphics[width=0.45\textwidth]{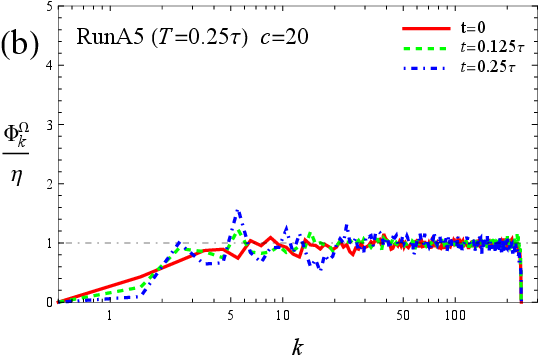}
\caption{Time development of enstrophy flux $\Phi_k^\Omega$ for states with quasi-constant enstrophy flux obtained in the numerical searches with time interval (a) $T=0$ (RunA0) and (b) $T=0.25\tau$ (RunA5).}
\label{fig:trnsfA}
\end{figure}
\begin{figure}[h!]
\includegraphics[width=0.45\textwidth]{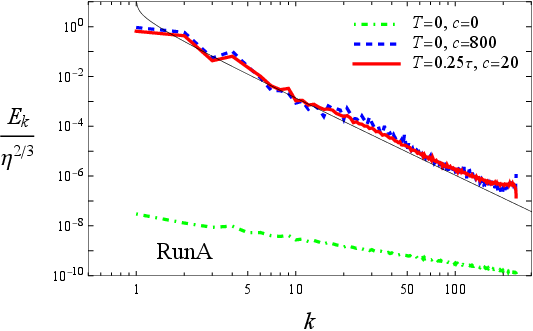}
\caption{Energy spectra $E_k(\vb*{\omega}_t)$ at time $t=0$ for states with quasi-constant enstrophy flux in time intervals $T=0$ (RunA0) and $T=0.25\tau$ (RunA5) obtained in the numerical searches. The thin solid line shows $E_k=C_K\eta^{2/3}k^{-3}(\ln(k/k_{\mathrm{b}}))^{-1/3}$ with $C_K=1.81$ (LRA), $\eta=1$ and $k_{\mathrm{b}}=1$.}
\label{fig:espeA}
\end{figure}

Since the enstrophy flux in wavenumber space $\Phi^\Omega_k(\vb*{\omega})$ is 
not a conserved quantity under inviscid motion, a state with constant enstrophy flux instantaneously does not maintain this property under time evolution according to the Euler equation in general.  
As shown in Fig.~\ref{fig:trnsfA} (a), a state $\vb*{\omega}$ 
numerically searched in RunA0 ($T=0$) approximates a state with 
constant enstrophy flux within the achieved precision instantaneously,
but $\Phi^\Omega_k(\vb*{\omega}(t))$ deviates from the constant $\eta$ 
as $t$ increases. On the other hand, a state in which the enstrophy
flux remains quasi-constant in wavenumber space over
the time interval $T=0.25\tau$ is successfully obtained in
RunA5 ($T=0.25\tau$), as shown in Fig.~\ref{fig:trnsfA} (b).  

Figure~\ref{fig:espeA} confirms the result of Y22 that 
the state developed through MC cycles ($c=800$) in RunA0 ($T=0$)
is consistent with the energy spectrum $E_k$ of the form 
Eq.~(\ref{eq:Ek_ens}). Since the enstrophy flux does not remain 
quasi-constant under time evolution for this state (Fig.~\ref{fig:trnsfA} (a)), the energy spectrum
also changes slightly with time, but the change
is small enough that $E_k$ at $t=0.25\tau$ almost overlaps with 
that at $t=0$ (plots omitted in Fig.~\ref{fig:espeA} for visibility).  
It is observed that the energy spectrum $E_k$ at $t=0$ changes 
moderately and stays close to Eq.~(\ref{eq:Ek_ens}) throughout
Series A. The energy spectrum at $t=0$ for the final state ($c=80$) in RunA5 ($T=0.25\tau$) is also shown in Fig.~\ref{fig:espeA}.  
Note that quasi-constant enstrophy flux $\Phi_k^\Omega$ in
$k_{\min}\le k \le k_{\max}$ and $0\le t \le T$ implies that 
the energy spectrum $E_k$ is quasi-stationary in the same
wavenumber range and time interval. 

\begin{figure}[t]
\includegraphics[width=0.45\textwidth]{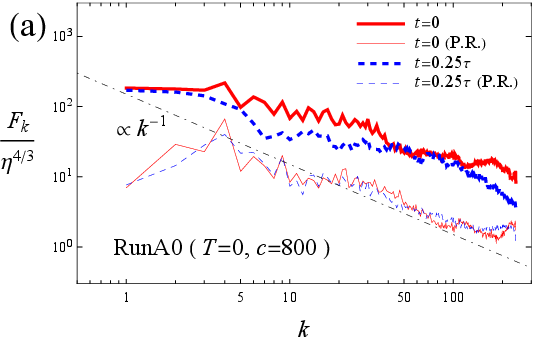}
\includegraphics[width=0.45\textwidth]{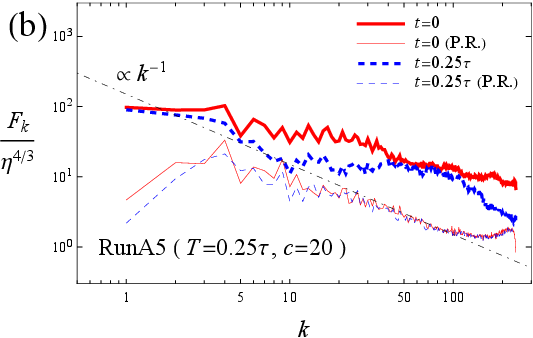}
\caption{The spectrum $F_k(\vb*{\omega}(t))$ of the field $(\omega(\vb*{x}))^2$ at time $t=0$ and $0.25\tau$ for states with quasi-constant enstrophy flux in time intervals (a) $T=0$ (RunA0) and (b) $T=0.25\tau$ (RunA5) obtained in the numerical searches.  Corresponding spectra for phase-randomized (P.R.) fields are also shown. The thin dot-dashed line represents the reference $F_k=150 \eta^{4/3}k^{-1}$. }
\label{fig:vor2speA}
\end{figure}
\begin{figure}[h!]
\includegraphics[width=0.45\textwidth]{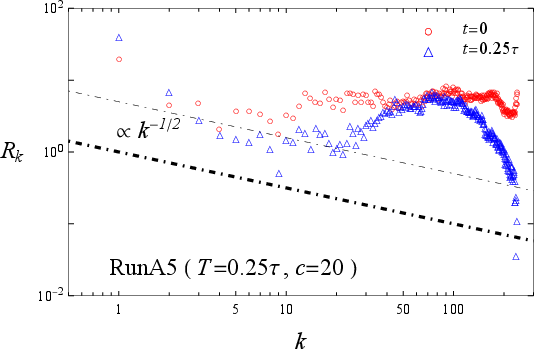}
\caption{Residual spectra $R_k(\vb*{\omega}(t))$ at time $t=0$ and $0.25\tau$ for states with quasi-constant enstrophy flux in time interval $T=0.25\tau$ (RunA5) obtained in the numerical searches.  The thick and thin dot-dashed lines represent $R_k=(k/\Delta k)^{-1/2}$ and $R_k=5 (k/\Delta k)^{-1/2}$, respectively.}
\label{fig:vor2ndspeA}
\end{figure}

\begin{figure}[t]
  \includegraphics[width=0.23\textwidth]{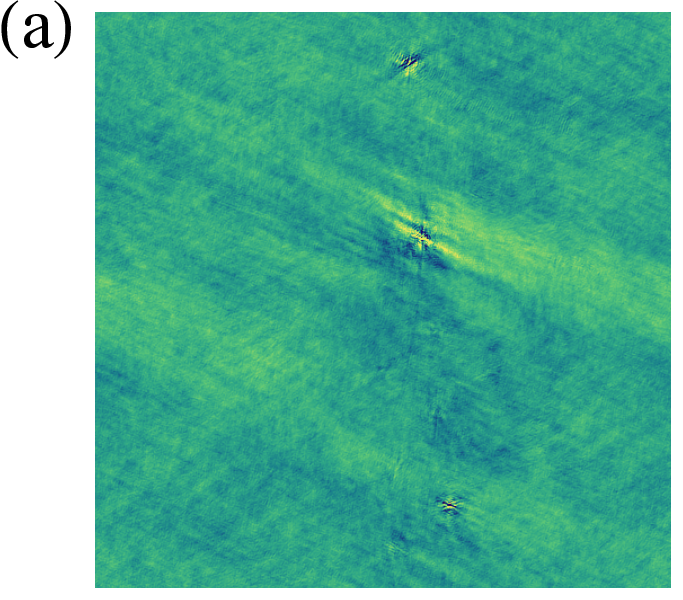}
  \includegraphics[width=0.23\textwidth]{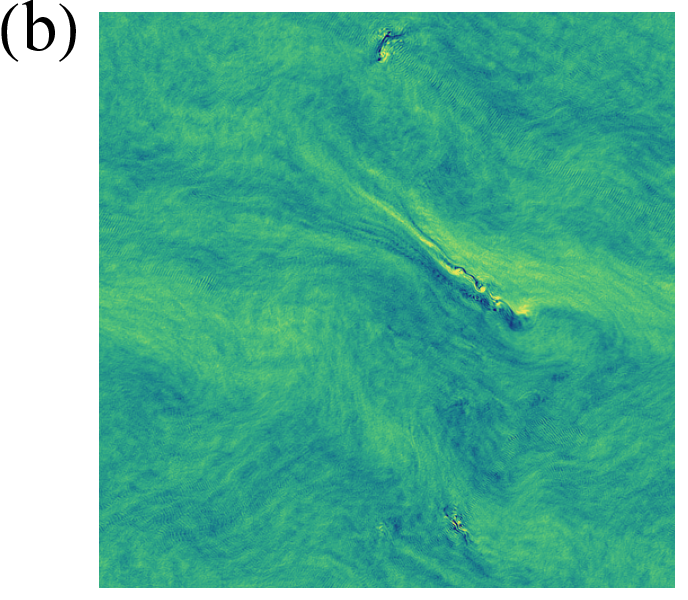}
  \caption{Vorticity fields in physical space at (a) $t=0$ and (b) $t=0.25\tau$ for the state with quasi-constant enstrophy flux in time interval $T=0.25\tau$ (RunA5). Bright (dark) regions denote positive (negative) vorticity.}
  \label{fig:vorxA}
\end{figure}

The energy spectrum $E_k$ and the enstrophy flux $\Phi^\Omega_k$
are second- and third-order quantities, respectively, in the velocity field 
$\vb*{u}_{\vb*{k}}$ or the vorticity field $\omega_{\vb*{k}}$.  
It is of interest to examine higher-order quantities.  
An example of the fourth-order quantity is the spectrum of $(\omega(\vb*{x}))^2$,
\begin{align}
F_k(\vb*{\omega})&:=(\Delta k)^{-1}
\sum_{\substack{\vb*{k}'\\(k-\Delta k/2 \le k'<k+\Delta k/2)}}^{\Delta}
\frac{1}{2}(\Delta k)^2|(\omega^2)_{\vb*{k}'}|^2,
\label{eq:F_k}
\end{align}
where $(\omega^2)_{\vb*{k}}$ is the Fourier transform of $(\omega(\vb*{x}))^2$.  
Define a phase-randomized state $\tilde{\vb*{\omega}}$ by 
$\tilde\omega_{\vb*{k}}=|\omega_{\vb*{k}}|\ee^{\ri\theta_{\vb*{k}}}$,
where $\theta_{\vb*{k}}$ are mutually independent
uniform random variables on $[0,2\pi)$ with
$\tilde{\omega}_{-\vb*{k}}=\tilde{\omega}_{\vb*{k}}^*$.  
This preserves second-order quantities but alters higher-order moments.  
The normalized residual spectrum is defined as
\begin{equation}
  R_k(\vb*{\omega}):=
  \frac{F_k(\vb*{\omega})-F_k(\tilde{\vb*{\omega}})}{F_k(\tilde{\vb*{\omega}})},\label{eq:residual}
\end{equation}
providing a measure of coherence in the fourth-order moments.

Figure~\ref{fig:vor2speA} (a) shows that $F_k(\vb*{\omega}(t))$ at $t=0$ for RunA0 is substantially larger than $F_k(\tilde{\vb*{\omega}}(t))$, indicating strong coherence. This coherence diminishes at $t=0.25\tau$, but remains higher than the typical turbulence level $F_k=150 \eta^{4/3}k^{-1}$ estimated from DNS (see also Fig.~\ref{fig:vor2speDS}(a) ).  For RunA5, as shown in Figure~\ref{fig:vor2speA} (b), the spectra $F_k(\vb*{\omega}(t))$ are smaller, implying coherence decreases as $T$ increases, approaching turbulence state values.
Note that the spectrum $F_k(\vb*{\omega}(t))$ varies significantly with time $t$ for the final state ($c=20$) in RunA5 although the energy spectrum $E_k(\vb*{\omega}(t))$ and the enstrophy flux $\Phi_k^\Omega(\vb*{\omega}(t))$ are quasi-stationary in the time interval $[0,T_5]$.

The degree of coherence can be seen more quantitatively by examing the normalized residual spectrum $R_k(\vb*{\omega}(t))$ as shown in Fig.~\ref{fig:vor2ndspeA} for RunA5.  The coherence in the state obtained in RunA5 remains considerably stronger compared to the turbulence state in DNS. Specifically, the spectra $R_k(\vb*{\omega}(t))$ for RunA5 are approximately at the level of $5 (k/\Delta k)^{-1/2}$ or larger across the entire wavenumber range, whereas $R_k(\vb*{\omega}(t))$ in DNS stays around the level of $(k/\Delta k)^{-1/2}$ and seldom exceeds $5 (k/\Delta k)^{-1/2}$ (see Fig.~\ref{fig:vor2speDS}(b)).  See Appendix 
for a possible interpretation of the scaling $R_k(\vb*{\omega})\propto(k/\Delta k)^{-1/2}$.  
It is also found that the spectra $R_k(\vb*{\omega}_t)$ of RunA4 ($T=0.125\tau$, $c=16$) and RunA5 ($T=0.25\tau$, $c=20$) almost coincide.  The plots for RunA4 in Fig.~\ref{fig:vor2ndspeA} are omitted for visibility. Thus, further decrease of $R_k$ with increasing time interval $T$ in the manner of RunA may be limited.

Figure~\ref{fig:vorxA} (a) shows vorticity fields for RunA5. Compared with RunA0 (Y22 Fig.3(a)), peak vorticity is reduced by a factor of 0.87, consistent with decreased $R_k$ and coherence. Self-advection leads to deformation from $t=0$ to $t=0.25\tau$ (see Fig.~\ref{fig:vorxA} (b)), and prior analysis indicate the advection reduces fourth-order coherence while maintaining quasi-constant enstrophy flux.

\begin{figure}[t]
\includegraphics[width=0.45\textwidth]{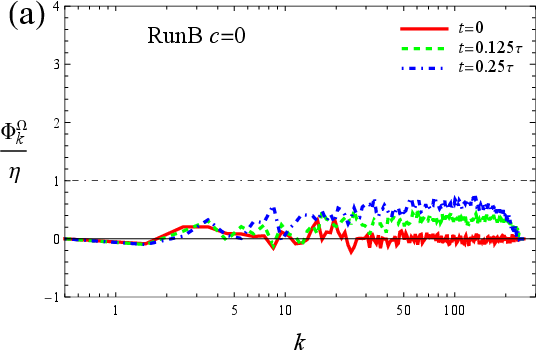}
\includegraphics[width=0.45\textwidth]{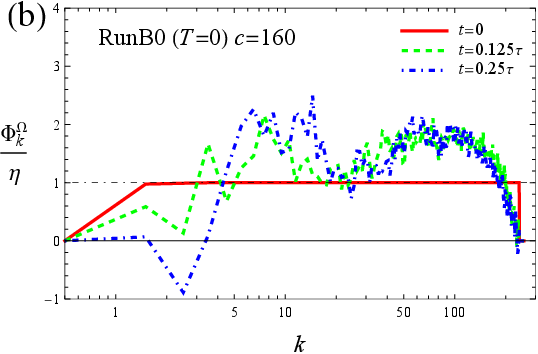}
\includegraphics[width=0.45\textwidth]{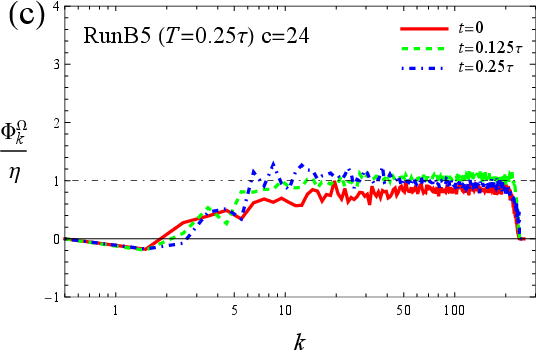}
\caption{Time development of enstrophy flux $\Phi_k^\Omega$ for states with quasi-constant enstrophy flux obtained in RunB: (a) initial ($T=0$), (b) RunB0 ($T=0$), and (c) RunB5 ($T=0.25\tau$).}
\label{fig:trnsfB}
\end{figure}
\begin{figure}[t]
\includegraphics[width=0.45\textwidth]{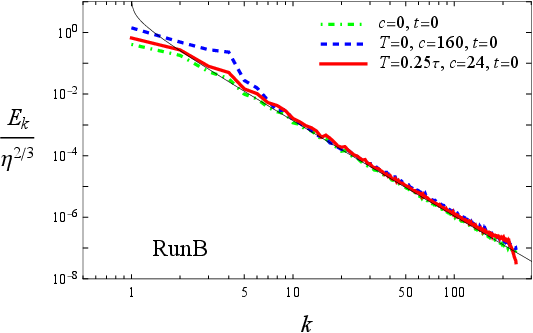}
\caption{Energy spectra $E_k$ at $t=0$ for the initial state (RunB $c=0$), RunB0 ($T=0$, $c=160$), and RunB5 ($T=0.25\tau$, $c=24$).}
\label{fig:espeB}
\end{figure}

In RunB, all initial states ($c=0$) are the same phase-randomized field with energy spectrum Eq.~(\ref{eq:Ek_ens}). Fig.~\ref{fig:trnsfB} (a) shows initial $\Phi_k^\Omega$ is near zero due to randomness, but positive flux emerges under inviscid dynamics. RunB0 yields the final state ($c=160$) with quasi-constant flux which varies moderately in time (Fig.~\ref{fig:trnsfB} (b)), in comparison with RunA0. RunB5 (Fig.~\ref{fig:trnsfB} (c)) yields a state with quasi-constant flux, but the acheived presion $\delta$ is about 3.6 times larger than RunA5 (Table~\ref{tab:runs}), indicating limited convergence within feasible MC cycles.

Figure~\ref{fig:espeB} shows energy spectra remain stable throughout RunB except for low wavenumber elevation, which is suppressed in RunB5. All energy spectra are nearly stationary under inviscid dynamics up to $t=0.25\tau$.

\begin{figure}[t]
\includegraphics[width=0.45\textwidth]{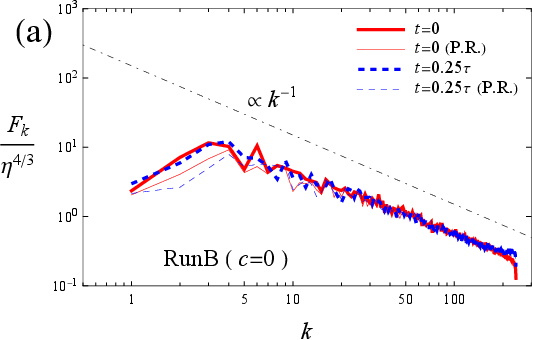}
\includegraphics[width=0.45\textwidth]{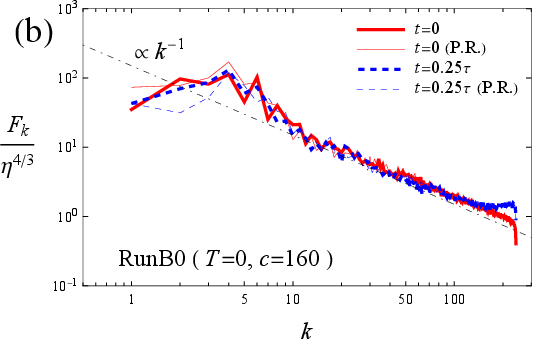}
\includegraphics[width=0.45\textwidth]{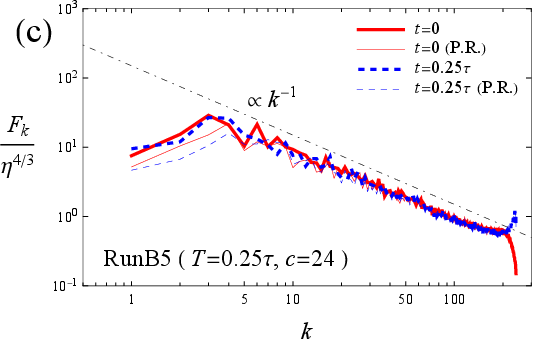}
\caption{Fourth-order spectra $F_k$ at $t=0$ and $t=0.25\tau$ for initial state (RunB $c=0$), RunB0 ($T=0$, $c=160$), and RunB5 ($T=0.25\tau$, $c=24$). Corresponding spectra for phase-randomized (P.R.) fields are also shown. The thin dot-dashed line represents the reference $F_k=150 \eta^{4/3}k^{-1}$.}  
\label{fig:vor2speB}
\end{figure}

\begin{figure}[t]
\includegraphics[width=0.45\textwidth]{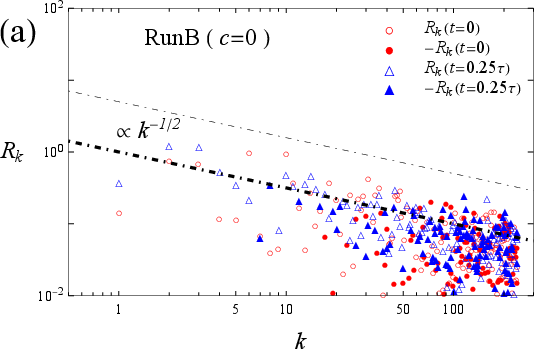}
\includegraphics[width=0.45\textwidth]{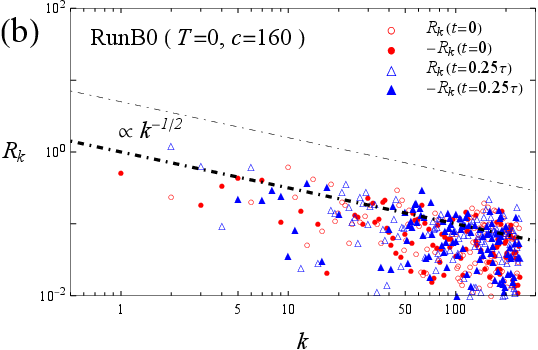}
\includegraphics[width=0.45\textwidth]{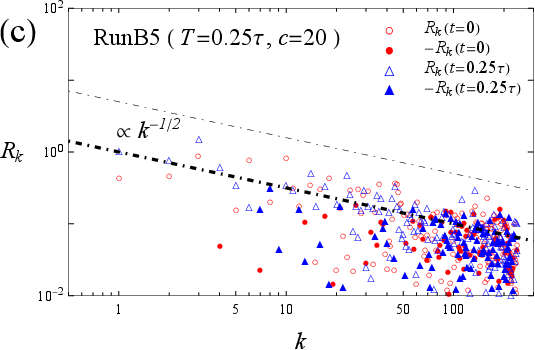}
\caption{Normalized residual spectra $R_k$ at $t=0$ and $0.25\tau$ for initial state (RunB $c=0$), RunB0 ($T=0$, $c=160$), and RunB5 ($T=0.25\tau$, $c=24$). The thick and thin dot-dashed lines represent $R_k=(k/\Delta k)^{-1/2}$ and $R_k=5 (k/\Delta k)^{-1/2}$, respectively.}
\label{fig:vor2ndspeB}
\end{figure}

The spectrum $F_k$ for the initial state ($c=0$) of RunB is substantially smaller than the reference level $F_k=150 \eta^{4/3}k^{-1}$ as shown in Fig.~\ref{fig:vor2speB} (a).  The spectra increase during RunB and show approximate $k^{-1}$ scaling with magnitudes largest for RunB0 for smallest for RunB5.  However, spectra $F_k(\vb*{\omega}(t))$ and their phase-randomized counterparts $F_k(\tilde{\vb*{\omega}}(t))$ nearly coincide, indicating little fourth-order coherence developed during RunB.  

We further analyse the coherence by investigating the residual spectra $R_k(\vb*{\omega}(t))$.  The spectra at $t=0$ and $0.25\tau$ are shown for the initial state of RunB ($c=0$) and the final states of RunB0 ($T=0$, $c=160$) and RunB5 ($T=0.25\tau$, $c=24$) in Fig.~\ref{fig:vor2ndspeB}. The sign of $R_k(\vb*{\omega}(t))$ varies irregularly with wavenumber $k$, and the magnitudes scatter around $(k/\Delta k)^{-1/2}$, not exceeding $5(k/\Delta k)^{-1/2}$.  We do not observe systematic changes of $R_k(\vb*{\omega}(t))$ during RunB, confirming the prior result that the coherence do not develop during RunB.  By comparing $R_k(\vb*{\omega}(t))$ with that in DNS (Fig.~\ref{fig:vor2speDS} (b)), we may conclude that the states with quasi-constant enstrophy flux obtained in RunB are less coherent in terms of their fourth-order moments than the turbulence state in DNS.  

\begin{figure}[t]
  \includegraphics[width=0.23\textwidth]{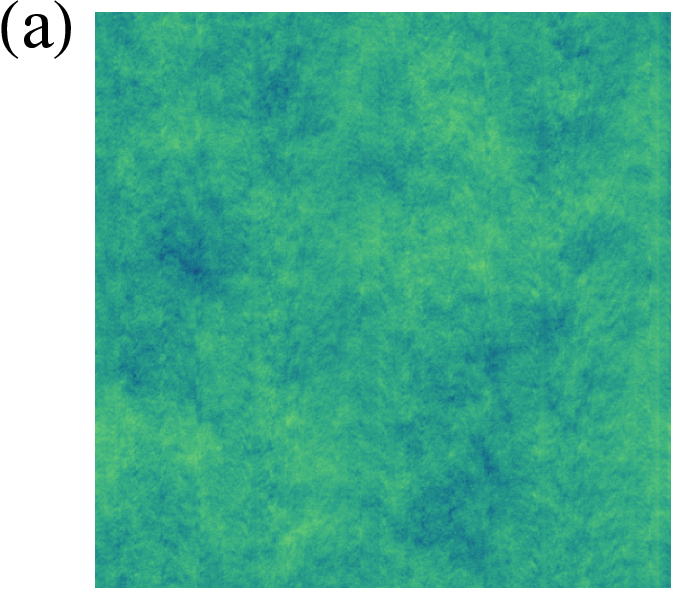}
  \includegraphics[width=0.23\textwidth]{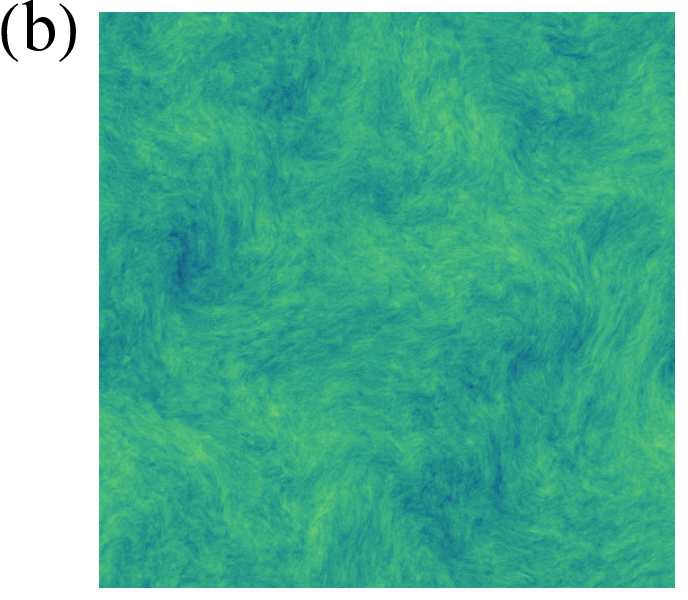}
  \includegraphics[width=0.23\textwidth]{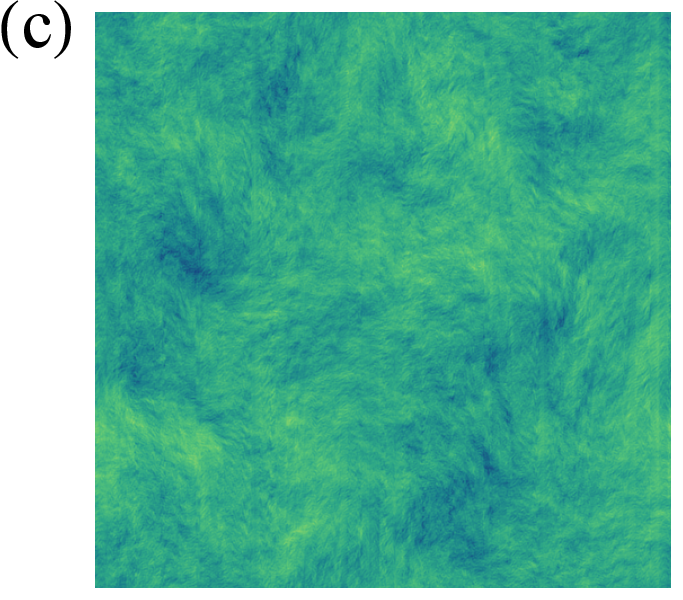}
  \includegraphics[width=0.23\textwidth]{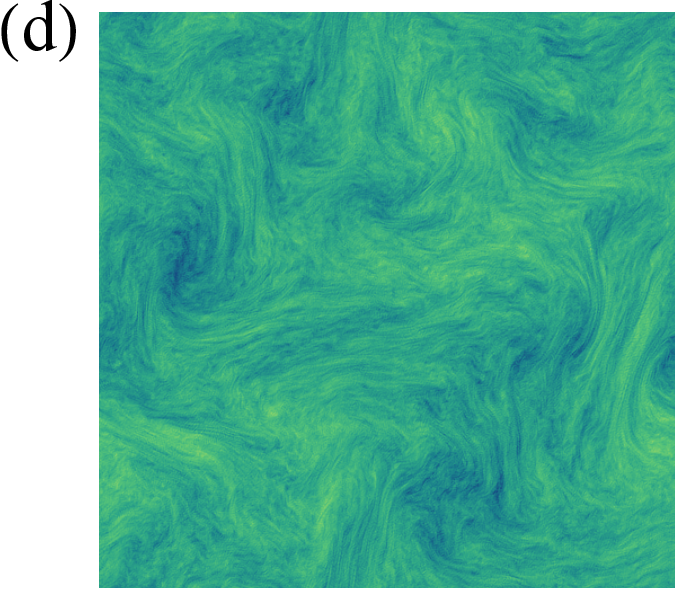}
\caption{Vorticity fields for RunB initial state ($c=0$) at (a) $t=0$ and (b) $t=0.25\tau$, and for RunB5 final state ($c=24$) at (c) $t=0$ and (d) $t=0.25\tau$.}
\label{fig:vorxB}
\end{figure}

The vorticity field in physical space for the initial state ($c=0$) of RunB at $t=0$ is shown in Fig.~\ref{fig:vorxB} (a).  The state shows mixed-scale structures reflecting Eq.~(\ref{eq:Ek_ens}).  This evolves over time into stretched structures via self-advection as shown in Fig.~\ref{fig:vorxB} (b) ($t=0.25\tau$).  It is remarkable that the emergence of the stretched structures is also observed during the numerical search for states with constant enstrophy flux in RunB5, as shown in the vorticity field of state in RunB5 at $t=0$ (Fig.~\ref{fig:vorxB} (c)).  The vorticity field is further stretched during the dynamical evolution, as shown in Fig.~\ref{fig:vorxB} (d) ($t=0.25\tau$).  

\section{Discussions}
\label{sec:discussions}

We attempted numerical searches for states with constant enstrophy flux in 
wavenumber space over time intervals up to $T=0.25\tau$
using a Monte Carlo method.
We performed two series of numerical searches, RunA and RunB, which differed in their settings.
The obtained states in the two series differed when
compared with respect to the coherence of the fourth-order moment, as measured by 
the normalized residual spectrum $R_k(\vb*{\omega})$. 

Ideally, the numerical searches should provide unbiased samples from 
the ensemble of states with constant enstrophy flux given by 
Eq.~(\ref{eq:Pfluxconst}). However, actual numerical searches 
face limitations. For example, the error function 
$\Delta^{(T)}(\vb*{\omega})$ does not reach zero strictly, and a finite tolerance
$\delta >0$, whose critical value is not known a priori, is inevitable. 
States can become trapped during the searches at local minima 
of $\Delta^{(T)}(\vb*{\omega})$ outside 
the target ensemble or in atypical states within the ensemble.  

The choice of initial states in the searches may also affect the quality of the resulting states.  The discrepancy in obtained states between RunA and RunB regarding $R_k(\vb*{\omega})$ is possibly explained by the limitations of either or both numerical searches. We find that the spectra $R_k(\vb*{\omega})$ in RunB are closer to those in DNS.  A possible interpretation of these results, favorable to the conjecture presented in Sec.~\ref{sec:introduction}, is as follows. The numerical searches of RunB were relatively successful in reaching typical states with constant flux whose statistical characteristics resemble those of turbulence states, whereas RunA yielded atypical states that do not resemble turbulence states in some statistical aspects.  The observation of stretched structures in the vorticity field of RunB (see Figs.~\ref{fig:vorxB} (c) and (d)) may provide additional support for RunB being relatively successful in approximating turbulence states.  By focusing on the result of RunA that the spectrum $R_k(\vb*{\omega)}$ decreases and approaches that of turbulence states as the time interval $T$ increases, we may further anticipete that $R_k(\vb*{\omega})$ in RunA will converge to that of turbulence states for very large $T$.  

Although stretched structures are observed in the vorticity fields of the numerically obtained quasi-constant-flux states, as shown in Figs.~{\ref{fig:vorxA}} and {\ref{fig:vorxB}}, their visual appearance may differ somewhat from that of the vorticity fields in DNS (see, e.g., Fig.~5(a) of Ref.~{\cite{Yoshida2022}}). The spectra $E_k(\vb*{\omega})$, $F_k(\vb*{\omega})$, and $R_k(\vb*{\omega})$ do not fully capture the differences observed in these images.
  It is desirable to develop quantitative metrics that can objectively characterize these differences. While we have focused on quantities in wavenumber space in this paper, quantities in physical space, such as the velocity structure functions $S_p(r)=\expval{(u_j(\vb*{x}+r \vb*{e}_j)- u_j(\vb*{x}))^p}$, where $u_j(\vb*{x})$ denotes the $j$th component of the velocity field and $\vb*{e}_j$ is the unit vector in the direction of $x_j$, may serve as promising candidates for future studies. The condition of constant enstrophy flux $\Phi^\Omega_k=\eta$ corresponds to the law $S_3(r)=(1/8)\eta r^3$ {\cite{Bernard1999}}, and our preliminary analysis confirms that this law approximately holds for the states with constant enstrophy flux obtained in the present numerical searches (figure omitted). Structure functions of higher order $S_p(r)$ $(p \ge 4)$ may contain information about intense vorticity structures.

In summary, 
the present results of
the numerical searches are insufficient to support the conjecture conclusively.  It is necessary to perform many more series of numerical searches
with various types of initial conditions to obtain
strong evidence supporting or rejecting the conjecture.  

The main obstacle to performing additional searches 
is computational cost and time.  
Computation of the dynamical evolution of the state over
the time interval $T$ is required at each Monte-Carlo
(MC)
 step, in which $\omega_{\vb*{k}}$ is altered for only a single wavenumber $\vb*{k}$.  
 One MC cycle consists of $\pi N^2/9$ MC steps. For example, RunB5 ($N=512$, $c=24$, $T=0.25\tau$) required the computation of dynamical evolution over a total time interval of about $5.5\times 10^5 \tau$, where $\tau:=\eta^{-1/3}$. This computational cost is enormous compared with the time interval of at most $10^2\tau$ typically required in DNS to obtain statistically quasi-stationary states that may represent turbulence states. 
The Monte Carlo method employed in the present study 
is primitive and unbiased but highly inefficient.
As seen in Fig.~\ref{fig:trnsfB} (a),
the dynamical evolution of the state itself can be
an efficient way to reach a state with constant flux
when a suitable initial state is chosen.
However, such a search is completely biased by
dynamical preferences and is not useful for verification of
the conjecture. To proceed with verifying
the conjecture, it may be necessary to develop a more efficient
and feasible method, allowing inevitable biases to some extent,
to search for states with constant flux.  
Finally, under the optimistic expectation that the conjecture will be resolved affirmatively, practical applications of numerical searches for constant-flux states are anticipated. At that stage, achieving efficiency comparable to that of DNS will become an important research task.

\section*{Acknowledgment}
This research used computational resources of Wisteria/BDEC-01 Odyssey (the University of Tokyo), provided by the Multidisciplinary Cooperative Research Program in the Center for Computational Sciences, University of Tsukuba.  This work was supported by JSPS KAKENHI Grant Number JP24K06880. 

\bibliographystyle{apsrev4-2}
\bibliography{myref}

\begin{thebibliography}{18}%
\makeatletter
\providecommand \@ifxundefined [1]{%
 \@ifx{#1\undefined}
}%
\providecommand \@ifnum [1]{%
 \ifnum #1\expandafter \@firstoftwo
 \else \expandafter \@secondoftwo
 \fi
}%
\providecommand \@ifx [1]{%
 \ifx #1\expandafter \@firstoftwo
 \else \expandafter \@secondoftwo
 \fi
}%
\providecommand \natexlab [1]{#1}%
\providecommand \enquote  [1]{``#1''}%
\providecommand \bibnamefont  [1]{#1}%
\providecommand \bibfnamefont [1]{#1}%
\providecommand \citenamefont [1]{#1}%
\providecommand \href@noop [0]{\@secondoftwo}%
\providecommand \href [0]{\begingroup \@sanitize@url \@href}%
\providecommand \@href[1]{\@@startlink{#1}\@@href}%
\providecommand \@@href[1]{\endgroup#1\@@endlink}%
\providecommand \@sanitize@url [0]{\catcode `\\12\catcode `\$12\catcode
  `\&12\catcode `\#12\catcode `\^12\catcode `\_12\catcode `\%12\relax}%
\providecommand \@@startlink[1]{}%
\providecommand \@@endlink[0]{}%
\providecommand \url  [0]{\begingroup\@sanitize@url \@url }%
\providecommand \@url [1]{\endgroup\@href {#1}{\urlprefix }}%
\providecommand \urlprefix  [0]{URL }%
\providecommand \Eprint [0]{\href }%
\providecommand \doibase [0]{https://doi.org/}%
\providecommand \selectlanguage [0]{\@gobble}%
\providecommand \bibinfo  [0]{\@secondoftwo}%
\providecommand \bibfield  [0]{\@secondoftwo}%
\providecommand \translation [1]{[#1]}%
\providecommand \BibitemOpen [0]{}%
\providecommand \bibitemStop [0]{}%
\providecommand \bibitemNoStop [0]{.\EOS\space}%
\providecommand \EOS [0]{\spacefactor3000\relax}%
\providecommand \BibitemShut  [1]{\csname bibitem#1\endcsname}%
\let\auto@bib@innerbib\@empty
\bibitem [{\citenamefont {Bedrossian}\ \emph {et~al.}(2022)\citenamefont
  {Bedrossian}, \citenamefont {Blumenthal},\ and\ \citenamefont
  {Punshon-Smith}}]{Bedrossian2022batchelor}%
  \BibitemOpen
  \bibfield  {author} {\bibinfo {author} {\bibfnamefont {J.}~\bibnamefont
  {Bedrossian}}, \bibinfo {author} {\bibfnamefont {A.}~\bibnamefont
  {Blumenthal}},\ and\ \bibinfo {author} {\bibfnamefont {S.}~\bibnamefont
  {Punshon-Smith}},\ }\href@noop {} {\bibfield  {journal} {\bibinfo  {journal}
  {Communications on Pure and Applied Mathematics}\ } (\bibinfo {year}
  {2022})},\ \bibinfo {note} {doi:
  https://doi.org/10.1002/cpa.22022}\BibitemShut {NoStop}%
\bibitem [{\citenamefont {Kawahara}\ and\ \citenamefont
  {Kida}(2001)}]{KawaharaKida2001}%
  \BibitemOpen
  \bibfield  {author} {\bibinfo {author} {\bibfnamefont {G.}~\bibnamefont
  {Kawahara}}\ and\ \bibinfo {author} {\bibfnamefont {S.}~\bibnamefont
  {Kida}},\ }\href@noop {} {\bibfield  {journal} {\bibinfo  {journal} {Jounal
  of Fluid Mechanics}\ }\textbf {\bibinfo {volume} {449}},\ \bibinfo {pages}
  {291} (\bibinfo {year} {2001})}\BibitemShut {NoStop}%
\bibitem [{\citenamefont {Martin}\ \emph {et~al.}(1973)\citenamefont {Martin},
  \citenamefont {Siggia},\ and\ \citenamefont {Rose}}]{MartinSiggiaRose1973}%
  \BibitemOpen
  \bibfield  {author} {\bibinfo {author} {\bibfnamefont {P.}~\bibnamefont
  {Martin}}, \bibinfo {author} {\bibfnamefont {E.}~\bibnamefont {Siggia}},\
  and\ \bibinfo {author} {\bibfnamefont {H.}~\bibnamefont {Rose}},\ }\href@noop
  {} {\bibfield  {journal} {\bibinfo  {journal} {Phys. Rev. A}\ }\textbf
  {\bibinfo {volume} {8}},\ \bibinfo {pages} {423} (\bibinfo {year}
  {1973})}\BibitemShut {NoStop}%
\bibitem [{\citenamefont {Janssen}(1976)}]{Janssen1976}%
  \BibitemOpen
  \bibfield  {author} {\bibinfo {author} {\bibfnamefont {H.-K.}\ \bibnamefont
  {Janssen}},\ }\href@noop {} {\bibfield  {journal} {\bibinfo  {journal}
  {Zeitschrift f\"ur Physik B}\ }\textbf {\bibinfo {volume} {23}},\ \bibinfo
  {pages} {377} (\bibinfo {year} {1976})}\BibitemShut {NoStop}%
\bibitem [{\citenamefont {de~Dominicis}(1976)}]{deDominicis1976}%
  \BibitemOpen
  \bibfield  {author} {\bibinfo {author} {\bibfnamefont {C.}~\bibnamefont
  {de~Dominicis}},\ }\href@noop {} {\bibfield  {journal} {\bibinfo  {journal}
  {Journal de Physique Colloques}\ }\textbf {\bibinfo {volume} {37}},\ \bibinfo
  {pages} {247} (\bibinfo {year} {1976})}\BibitemShut {NoStop}%
\bibitem [{\citenamefont {Canet}\ \emph {et~al.}(2016)\citenamefont {Canet},
  \citenamefont {Delamotte},\ and\ \citenamefont
  {Wschebor}}]{CanetDelamotteWschebor2016}%
  \BibitemOpen
  \bibfield  {author} {\bibinfo {author} {\bibfnamefont {L.}~\bibnamefont
  {Canet}}, \bibinfo {author} {\bibfnamefont {B.}~\bibnamefont {Delamotte}},\
  and\ \bibinfo {author} {\bibfnamefont {N.}~\bibnamefont {Wschebor}},\
  }\href@noop {} {\bibfield  {journal} {\bibinfo  {journal} {Physcal Review E}\
  }\textbf {\bibinfo {volume} {93}},\ \bibinfo {pages} {063101} (\bibinfo
  {year} {2016})}\BibitemShut {NoStop}%
\bibitem [{\citenamefont {Kraichnan}(1965)}]{Kraichnan1965}%
  \BibitemOpen
  \bibfield  {author} {\bibinfo {author} {\bibfnamefont {R.~H.}\ \bibnamefont
  {Kraichnan}},\ }\href@noop {} {\bibfield  {journal} {\bibinfo  {journal}
  {Phys. Fluids}\ }\textbf {\bibinfo {volume} {8}},\ \bibinfo {pages} {575}
  (\bibinfo {year} {1965})}\BibitemShut {NoStop}%
\bibitem [{\citenamefont {Kaneda}(1981)}]{Kaneda1981}%
  \BibitemOpen
  \bibfield  {author} {\bibinfo {author} {\bibfnamefont {Y.}~\bibnamefont
  {Kaneda}},\ }\href@noop {} {\bibfield  {journal} {\bibinfo  {journal} {J.
  Fluid Mech.}\ }\textbf {\bibinfo {volume} {107}},\ \bibinfo {pages} {131}
  (\bibinfo {year} {1981})}\BibitemShut {NoStop}%
\bibitem [{\citenamefont {Zhou}(2021)}]{Zhou2021}%
  \BibitemOpen
  \bibfield  {author} {\bibinfo {author} {\bibfnamefont {Y.}~\bibnamefont
  {Zhou}},\ }\href
  {https://doi.org/https://doi.org/10.1016/j.physrep.2021.07.001} {\bibfield
  {journal} {\bibinfo  {journal} {Physics Reports}\ }\textbf {\bibinfo {volume}
  {935}},\ \bibinfo {pages} {1} (\bibinfo {year} {2021})}\BibitemShut {NoStop}%
\bibitem [{\citenamefont {Yoshida}(2022)}]{Yoshida2022}%
  \BibitemOpen
  \bibfield  {author} {\bibinfo {author} {\bibfnamefont {K.}~\bibnamefont
  {Yoshida}},\ }\href@noop {} {\bibfield  {journal} {\bibinfo  {journal} {Phys.
  Rev. E}\ }\textbf {\bibinfo {volume} {106}},\ \bibinfo {pages} {045106}
  (\bibinfo {year} {2022})}\BibitemShut {NoStop}%
\bibitem [{\citenamefont {Richardson}(1922)}]{inRichardson1922}%
  \BibitemOpen
  \bibfield  {author} {\bibinfo {author} {\bibfnamefont {L.~F.}\ \bibnamefont
  {Richardson}},\ }\bibinfo {title} {Weather prediction by numerical
  processes}\ (\bibinfo  {publisher} {Cambridge University Press},\ \bibinfo
  {year} {1922})\ p.~\bibinfo {pages} {66}\BibitemShut {NoStop}%
\bibitem [{\citenamefont {Kolmogorov}(1941)}]{Kolmogorov1941a}%
  \BibitemOpen
  \bibfield  {author} {\bibinfo {author} {\bibfnamefont {A.~N.}\ \bibnamefont
  {Kolmogorov}},\ }\href@noop {} {\bibfield  {journal} {\bibinfo  {journal}
  {Dokl. Akad. Nauk SSSR}\ }\textbf {\bibinfo {volume} {30}},\ \bibinfo {pages}
  {301} (\bibinfo {year} {1941})},\ \bibinfo {note} {(reprinted in {\it Proc.
  R. Soc. Lond.} A 434:9-13)}\BibitemShut {NoStop}%
\bibitem [{\citenamefont {Bos}\ and\ \citenamefont
  {Rubinstien}(2017)}]{BosRubinstein2017}%
  \BibitemOpen
  \bibfield  {author} {\bibinfo {author} {\bibfnamefont {W.~J.~T.}\
  \bibnamefont {Bos}}\ and\ \bibinfo {author} {\bibfnamefont {R.}~\bibnamefont
  {Rubinstien}},\ }\href@noop {} {\bibfield  {journal} {\bibinfo  {journal}
  {Physical Review Fluid}\ }\textbf {\bibinfo {volume} {2}},\ \bibinfo {pages}
  {022601(R)} (\bibinfo {year} {2017})}\BibitemShut {NoStop}%
\bibitem [{\citenamefont {Ishihara}\ \emph {et~al.}(2016)\citenamefont
  {Ishihara}, \citenamefont {Morishita}, \citenamefont {Yokokawa},
  \citenamefont {Uno},\ and\ \citenamefont {Kaneda}}]{Ishiharaetal2016}%
  \BibitemOpen
  \bibfield  {author} {\bibinfo {author} {\bibfnamefont {T.}~\bibnamefont
  {Ishihara}}, \bibinfo {author} {\bibfnamefont {K.}~\bibnamefont {Morishita}},
  \bibinfo {author} {\bibfnamefont {M.}~\bibnamefont {Yokokawa}}, \bibinfo
  {author} {\bibfnamefont {A.}~\bibnamefont {Uno}},\ and\ \bibinfo {author}
  {\bibfnamefont {Y.}~\bibnamefont {Kaneda}},\ }\href@noop {} {\bibfield
  {journal} {\bibinfo  {journal} {Physical Review Fluids}\ }\textbf {\bibinfo
  {volume} {1}},\ \bibinfo {pages} {082403(R)} (\bibinfo {year}
  {2016})}\BibitemShut {NoStop}%
\bibitem [{\citenamefont {Kaneda}(1987)}]{Kaneda1987}%
  \BibitemOpen
  \bibfield  {author} {\bibinfo {author} {\bibfnamefont {Y.}~\bibnamefont
  {Kaneda}},\ }\href@noop {} {\bibfield  {journal} {\bibinfo  {journal} {Phys.
  Fluids}\ }\textbf {\bibinfo {volume} {30}},\ \bibinfo {pages} {2672}
  (\bibinfo {year} {1987})}\BibitemShut {NoStop}%
\bibitem [{\citenamefont {Kaneda}(2007)}]{Kaneda2007}%
  \BibitemOpen
  \bibfield  {author} {\bibinfo {author} {\bibfnamefont {Y.}~\bibnamefont
  {Kaneda}},\ }\href@noop {} {\bibfield  {journal} {\bibinfo  {journal} {Fluid
  Dynamics Research}\ }\textbf {\bibinfo {volume} {39}},\ \bibinfo {pages}
  {526} (\bibinfo {year} {2007})}\BibitemShut {NoStop}%
\bibitem [{\citenamefont {Ishihara}\ and\ \citenamefont
  {Kaneda}(2001)}]{IshiharaKaneda2001}%
  \BibitemOpen
  \bibfield  {author} {\bibinfo {author} {\bibfnamefont {T.}~\bibnamefont
  {Ishihara}}\ and\ \bibinfo {author} {\bibfnamefont {Y.}~\bibnamefont
  {Kaneda}},\ }\href@noop {} {\bibfield  {journal} {\bibinfo  {journal}
  {Physics of Fluids}\ }\textbf {\bibinfo {volume} {13}},\ \bibinfo {pages}
  {544} (\bibinfo {year} {2001})}\BibitemShut {NoStop}%
\bibitem [{\citenamefont {Bernard}(1999)}]{Bernard1999}%
  \BibitemOpen
  \bibfield  {author} {\bibinfo {author} {\bibfnamefont {D.}~\bibnamefont
  {Bernard}},\ }\href@noop {} {\bibfield  {journal} {\bibinfo  {journal}
  {Physical Review E}\ }\textbf {\bibinfo {volume} {60}},\ \bibinfo {pages}
  {6184} (\bibinfo {year} {1999})}\BibitemShut {NoStop}%
\end{thebibliography}%

\appendix*

\section{Direct Numerical simulations of two-dimensional hyperviscous Navier–Stokes equation}
\label{sec:DNS}

Here we present results from DNS of a two-dimensional hyperviscous incompressible fluid with random forcing in a periodic boundary box. The detailed settings of the simulation are given in the Appendix of Y22. The enstrophy flux $\Phi_k^{\Omega}(\vb*{\omega}(t))$ and the energy spectrum $E_k(\vb*{\omega}(t))$ averaged over the time interval $64 \le t \le 128$ are also provided in the Appendix of Y22. It can be confirmed that $\Phi_k^{\Omega}$ is quasi-constant in the wavenumber range $26 \le k \le 1100$, identified as the inertial range. The energy spectrum $E_k^{\Omega}$ in this range can be fitted to the scaling form Eq.~(\ref{eq:Ek_ens}) with $C_K=2.20$ and $k_{\mathrm{b}}=4.0$.

\begin{figure}[ht]
  \includegraphics[width=0.45\textwidth]{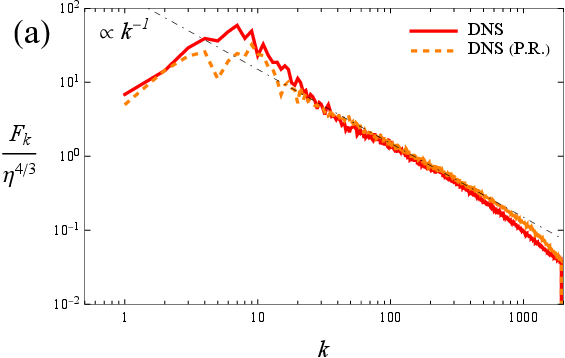}
  \includegraphics[width=0.45\textwidth]{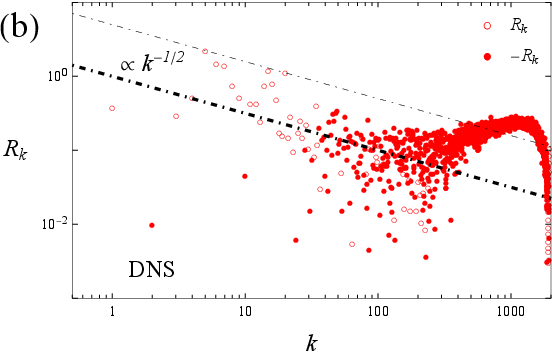}
\caption{(a) The spectrum $F_k(\vb*{\omega})$ of the field $(\omega(\vb*{x}))^2$ for the state $\vb*{\omega}$ obtained in DNS (solid line). The same spectrum $F_k(\tilde{\vb*{\omega}})$ for the phase-randomized state $\tilde{\vb*{\omega}}$ (dashed line). The thin dot-dashed line represents $F_k=150\,\eta^{4/3} k^{-1}$. (b) The normalized residual spectrum $R_k(\vb*{\omega})$ for the same state. The thick and thin dot-dashed lines represent $R_k=(k/\Delta k)^{-1/2}$ and $R_k=5 (k/\Delta k)^{-1/2}$, respectively.}  
\label{fig:vor2speDS}
\end{figure}

The spectrum $F_k(\vb*{\omega})$ of the field $(\omega(\vb*{x}))^2$, defined 
by Eq.~(\ref{eq:F_k}), is shown for the state at $t=128$ in
Fig.~\ref{fig:vor2speDS} (a). The spectrum is consistent with the scaling law
$F_k \sim \eta^{4/3} k^{-1}$ obtained from dimensional analysis, where
$F_k$ depends only on $\eta$ and $k$. The corresponding spectrum
$F_k(\tilde{\vb*{\omega}})$ for the phase-randomized state
$\tilde{\vb*{\omega}}$ is also shown.  
The two spectra approximately coincide in the inertial range, 
indicating that the coherence regarding
the fourth-order moments of the vorticity field is small.

Let $\mathcal{S}_k$ be the set of wavevectors $\vb*{k}'(\in \mathcal{K}^+)$
in the shell $k-\Delta k/2 \le k' < k+\Delta k/2$, and let $N_k$ be
the number of elements in $\mathcal{S}_k$.  
If $(\omega^2)_{\vb*{k}'} - (\tilde{\omega}^2)_{\vb*{k}'}$ for 
$\vb*{k}' \in \mathcal{S}_k$ are mutually independent random variables 
with mean zero and standard deviation proportional to
$F_k(\vb*{\omega})/N_k$, then the standard deviation of
the normalized residual spectrum $R_k(\vb*{\omega})$ defined
by Eq.~(\ref{eq:residual}) scales as $N_k^{-1/2}$.
Since $N_k \sim \pi k/\Delta k$,
we have $|R_k(\vb*{\omega})| \propto (k/\Delta k)^{-1/2}$.  
Thus, even if 
$(\omega^2)_{\vb*{k}'} - (\tilde{\omega}^2)_{\vb*{k}'}$ has zero
mean, we may observe $|R_k(\vb*{\omega})| \propto (k/\Delta k)^{-1/2}$
due to the finite-size effect of $N_k$.  

The spectrum $R_k(\vb*{\omega})$ for the state at $t=128$ in
DNS is shown in Fig.~\ref{fig:vor2speDS} (b).  
The magnitude scatters around
$|R_k(\vb*{\omega})|\sim (k/\Delta k)^{-1/2}$ for $k$ in
the inertial range. This scaling can be attributed to the finite-size
effect of $N_k$, without implying genuine coherence in the state. 
For wavenumbers smaller than those in the inertial range, 
$R_k(\vb*{\omega})$ tends to be positive definite
and deviates from the scaling $(k/\Delta k)^{-1/2}$.
This may signal some coherence in the fourth-order moments. 
However, the deviations are not large, as
$R_k(\vb*{\omega})$ seldom exceeds $5(k/\Delta k)^{-1/2}$.

\end{document}